\documentclass[12pt]{iopart}
\usepackage{graphicx,iopams,euscript,setstack}
\DeclareMathAlphabet{\mathpzc}{OT1}{pzc}{mb}{it}
\newcounter{aaa}
\newcounter{bbb}
\newcommand{\ftext}[1]{\mbox{\textrm #1}}
\newcommand{\eqref}[1]{(\ref{#1})}
\newenvironment{subequations}{\refstepcounter{equation}
\setcounter{bbb}{\value{equation}}\setcounter{equation}{0}
\def\theequation{\arabic{bbb}\alph{equation}}}
{\setcounter{equation}{\value{bbb}}}%
\newenvironment{teor}[2][{}]{\begin{trivlist}\refstepcounter{aaa}%
\labelsep=0pt\item[\bfseries\itshape \small \theaaa. #2. ]#1}%
{\end{trivlist}}
\newenvironment{teor*}[2][{}]{\begin{trivlist}%
\labelsep=0pt\item[\bfseries\itshape \small #2 ]#1}%
{\end{trivlist}}
\newcommand{\ssy}[5]{#1  #4 \emph{#2} {\bf #3} #5}
\newcommand{\ogr}[2]{#1\,\vrule \,{}_{\displaystyle{}_{#2}}}

\newenvironment{proof}[1][.
]{\par\noindent\emph{Proof#1}}{\par\nopagebreak
\hfill$\square$\par}
\newcommand{\karti}[4]{\begin{figure}[#1]\begin{center}
\includegraphics[height=#2]{#3}
\end{center}\caption{#4}\end{figure}}
\newcommand{\lr}[2][{}]{ {<}#2{>}^{\mathstrut}_{\!#1}}
\newcommand{\lrj}[2][{}]{ {\leqslant} #2
{\geqslant}^{\mathstrut}_{#1}}
\newcommand{\varg}{\mathord{\mbox{\usefont{U}{pxmia}{m}{it}\selectfont
\char"31}}}
\newcommand{\R}{{\usefont{U}{pxsyb}{m}{n}\selectfont R}}
\newcommand{\Z}[1]{\mathcal V(#1)}
\newcommand{\Bd}{\mathop{\mathrm{Bd}}}
\newcommand{\lz}{\overset{_{\mathord\sqcap}}{\lambda}}
\newcommand{\A}{\ensuremath{M_{\mathcal N}}}
\newcommand{\M}{\ensuremath{M_\lozenge}}
\newcommand{\K}{\ensuremath{K_\lozenge}}
\newcommand{\Hu}{ \ensuremath{H_{\wedge}} }
\newcommand{\Hl}{\ensuremath{H_{\vee}} }
\newcommand{\Ou}{ \ensuremath{H'_{\wedge}} }
\newcommand{\Ol}{\ensuremath{H'_{\vee}} }
\newcommand{\Bu}{\EuScript H'_{\wedge}}
\newcommand{\Bl}{\EuScript H'_{\vee}}
\newcommand{\Blu}{\EuScript H'_{\wedge,\vee}}
\newcommand{\Out}{\tilde H'_{\wedge}}
\newcommand{\Olt}{ \tilde H'_{\vee}}

\newcommand{\Ola}{\ensuremath{H'_{{\mathcal N}\wedge}} }
\newcommand{\zv}[1]{#1^{\rm max}}
\newcommand{\C}{$\mathpzc{C}$}
\begin{document}
\title{No time machines in classical general relativity.}
\author{S Krasnikov}%
\address{Central Astronomical Observatory at Pulkovo,
St.Petersburg, 196140, Russia}
\ead{redish@pulkovo.spb.su}
\date{}
\begin{abstract}
Irrespective of local conditions imposed on the metric, any extendible
spacetime $U$ has a maximal extension containing no closed causal
curves outside the chronological past of $U$. We prove this fact and
interpret it as impossibility (in classical general relativity) of the
time machines, insofar as the latter are defined to be
causality-violating regions created by human beings (as opposed to
those appearing spontaneously).
\end{abstract}
\maketitle
\section{Introduction}

In this paper we prove a theorem which, in physical terms, says that
\emph{in classical general relativity a time machine cannot be
built}. To formulate the theorem and to substantiate such its
interpretation  we need some preliminary discussion.
\par
Suppose  one wants to undertake a time trip. A possible strategy would
be just to look for a ready-made closed timelike curve (CTC), or to
wait passively until such a curve appears.
\begin{teor}{Remark}\label{rem:DP}
It should be stressed that such expectation is not hopeless however
innocent the spacetime looks at the moment.  The possibility of a
`sudden', `unprovoked' appearance of a CTC, in my view, must be taken
quite seriously, neither theoretical, nor observational evidence
against them being known. Consider, for example, the Deutsch-Politzer
(DP) space \cite{politzer}, which is the spacetime obtained from the
Minkowski plane by making cuts along the segments $\{t=\pm1,
\;-1\leqslant x \leqslant1\}$ and  gluing then the upper bank of each cut
to the lower bank of the other cut (see figure~\ref{fig:dp}a).
\karti{h,b}{0.4\textwidth}{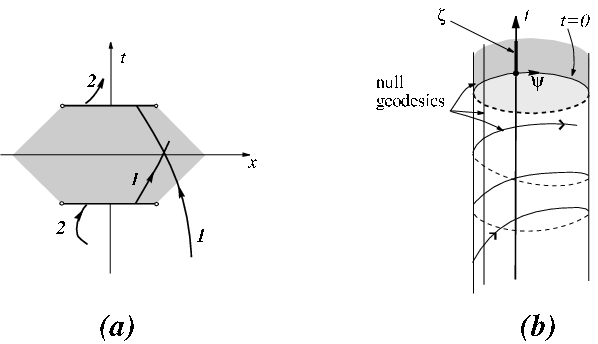}{\label{fig:dp} (a) The DP space. Curves
1 and 2 are actually continuous. Through each point of the shadowed
region pass closed and self-intersecting causal curves (such as 1).
(b) The extension $C$ of the Misner space. Causality is violated in
the shadowed region.} An observer located, say, at a point $x=0$,
$t=-5$ and fully informed about the geometry of the world at $t<-5$
cannot foretell whether the spacetime will evolve in the Minkowski
plane (preserving thus causality) or in the DP space, both, in
particular, being (in the four-dimensional case) the solutions of the
Einstein equations with the same (zero) source.
\end{teor}
Still, the
\emph{discovery} of a ready-made CTC is a matter of luck. The
alternative  would be the
\emph{creation} of such a curve.  In
particular, manipulating with matter
--- and thus according to the Einstein equations with the metric ---
an advanced civilization could try to \emph{force} the spacetime to
evolve into a time machine. It is the conjecture that, given a
suitable opportunity, the civilization can succeed \cite{MTY}, that
initiated intensive studies of the time machine (see \cite{revs} for
reviews and references). The difference between `finding' a closed
causal curve (or `causal loop' for brevity) $\ell$  and
`manufacturing' it is the central point of our consideration. In
distinguishing these two possibilities, I proceed from the
idea\footnote{For other proposed criteria see
\cite{Cr}.} that for the existence of $\ell$ to be attributable
to the activity that took place in a region $U$, $\ell$
and $U$ must satisfy at least the following two conditions:
\begin{enumerate}
  \item $\ell$ lies in the future of $U$, and not in its past;\label{it:fut}
  \item a  causal loop satisfying \eqref{it:fut} exists
  in\label{it:all} \emph{any allowed maximal} extension of $U$.
\end{enumerate}
Correspondingly, if no such $U$ can be found I consider $\ell$ as
`spontaneous' rather than  `manufactured' by anybody.
\par
The first of the conditions is self-evident, but the emphasized words
in \eqref{it:all} need some comment. An extendible spacetime $U$
typically has infinitely many maximal extensions (e.~g.\ if $U$ is the
Minkowski half-plane $t<0$ those are the Minkowski plane, the DP
space, any plane with the metric $\rmd s^2=(1+\omega^2)(\rmd x^2 -\rmd
t^2)$, where $\omega(t<0)=0$, etc.). However, they are not equipollent
from the  point of view of general relativity, since in that theory
often only those spacetimes are considered as possible models of the
Universe which
 satisfy some local conditions.
\begin{teor}{Definition}
We call a condition \C\ \emph{local} if the following is true:
\C\  holds in a spacetime $M$ if and only if it holds in any $U$
which is isometric to an open subset of $M$.
\end{teor}
Ideally --- that is, if we could describe all existing matter by a
single Lagrangian --- the local condition required by relativity would
be the  Einstein equations with the stress-energy tensor in the right
hand side corresponding to that Lagrangian. Of course we do not know
the Lagrangian and so in practice different more simple local
conditions (based, in particular, on our guesses about its properties)
are imposed. Examples are
\[
 G_{\alpha\beta}=-\Lambda \varg_{\alpha\beta},\quad\ftext{ or}
 \qquad G_{\alpha\beta}t^\alpha t^\beta\geqslant 0
\quad  \ftext{for any timelike } \bi t,
\]
where $G$ is the Einstein tensor and $\varg$ is the metric. To take
into account the r\^ole played by local conditions  I introduce the
notion of `\C-spacetime'.
\begin{teor}{Definition}\label{def:sp}
A smooth connected paracompact Hausdorff orientable manifold
endowed with a Lorentzian metric is a \emph{\C-spacetime} if it satisfies
a local condition \C.
\end{teor}
Correspondingly,
\begin{teor}{Definition}\label{def:ext}
A  \C-spacetime $M'$ is a \emph{\C-extension of} a  \C-spacetime $M$
if  $M$ is isometric to an open proper subset of $M'$. $M$ is called
\emph{\C-extendible} if it has a \C-extension and \emph{\C-maximal} otherwise.
\end{teor}
 Throughout the paper the letter
\C\ denotes the \emph{same} local condition.  I shall not
specify it though, because it is important that all the results below are
valid for \emph{any}
\C\ including trivial.
Obviously, in this latter case (when \C\ is trivial, that is when, in
fact, no additional local condition is imposed on the metric)
`\C-spacetime' and `\C-extension' are the same as `spacetime' and
`extension', respectively.
\par
Now we are in position to refine item~\eqref{it:all} in the above-formulated
criterion: the words `allowed maximal extension' stand
there for `\C-maximal \C-extension'.
\par
Imposing a local condition we still cannot provide uniqueness of
evolution of a spacetime. Whenever $M$ has a \C-maximal \C-extension
$M'$ (except when $M'=\overline{M}$) it has infinitely many other such
extensions \cite{chru,par} including, for example, those obtained from
$M'$ in the way we built the DP space from the Minkowski plane. I
interpret this as impossibility of creating a prescribed spacetime:
 whatever initial data (i.~e.\ the geometry
of  the  `initial  region' $M$) are prepared, and whatever are the equations
of motion of the matter  filling $M$, one does not know whether $M$
will evolve in the desired $M'$, or in any of other possible
extensions satisfying the same local conditions. However, in
 building a time machine it does not matter
 \emph{how exactly} the spacetime will evolve\footnote{Actually,
 the fact that one cannot predict the
evolution of a given spacetime is beneficial in building a time
machine being the protection against the time travel paradoxes
\cite{par}.}, but only whether
\emph{a} CTC will appear. So it would suffice to create a situation in
which a CTC is present in \emph{any} of the possible extensions (as is
sometimes the case with singularities \cite{HawEl}: fulfillment of
some local conditions guarantees the existence of a singularity in the
 extensions of some spacetimes even without fixing uniquely
 their evolution). On the other hand, \eqref{it:all} is
also a \emph{necessary} condition for considering a CTC as artificial.
In a theory (quantum gravity?) where different
\emph{probabilities} could be ascribed to  extensions,
it would not be the case. One could take credit for creating $\ell$
even if one's activity did not lead to its inevitable appearance, but
just increased its probability. However, any further discussion of
this hypothetical theory is beyond the scope of the present
paper.

\begin{teor*}{Theorem.} Any \C-spacetime $U$ has a \C-maximal  \C-extension
$\zv M$ such that all closed causal curves in $\zv M$ (if they exist
there) are confined to the chronological past of $U$.
\end{teor*}

Summing up we can say, that the theorem does not exclude at all the
possibility of a closed causal curve. But it shows that in
manufacturing such a curve it does not matter whether one, say, moves
the mouths of a wormhole, or just utters: `Abracadabra!'. The results
will be exactly the same: the curve may appear and it may not as well.
\begin{teor}{Example}
Consider a cylinder $C\equiv\{t\in{}$\R$^1,\:\psi=\psi+2\pi\}$ with
the metric $\varg\colon\quad\rmd s^2=-2\,\rmd\psi\rmd t -t\,\rmd
\psi^2$. The   part $U\equiv\{p\in C\colon\:t(p)<0\}$ of this cylinder
 is a causal spacetime called the Misner
space \cite{HawEl}, while the region $t\geqslant 0$ contains causal
loops. To see whether these loops are  an inevitable consequence of
something that takes place in the Misner space consider all possible
maximal extensions of the latter. Some of the extensions [e.~g.\
$(C,\varg)$] are acausal and some are not. Pick, for example, a
function $\Omega$ such that
\[\ogr{\Omega}{U}=1,\qquad \Omega(p)=0\quad\Leftrightarrow
\quad p\in\zeta,
\]
where $\zeta$ is the ray $\psi=0,\;t\geqslant 0$ (see
figure~\ref{fig:dp}b). The spacetime $(C',\Omega^{-1}\varg)$, where
$C'\equiv C-\zeta$, is a causal and, most likely,  maximal
extension of $U$. However, not all of these  extensions are equally
relevant. As is discussed above, we can impose a local condition \C\
and declare all spacetimes that do not obey it `unphysical'. For
example, we could require that a spacetime should be flat ($U$
is flat). Then $(C,\varg)$ would be an `allowed' extension (a
\C-extension) of $U$, while $(C',\Omega^{-1}\varg)$ would be not.
What the theorem asserts in application to this case is: whatever \C\
is chosen --- as long as  $(C,\varg)$ obeys it --- there can be found
a \emph{causal} maximal extension $\zv{M}$ of $U$ also obeying \C.
Let, for example, $\{M_n\}$,  $n=\dots,-1,0,1\dots$ be a set of
spacetimes each isometric to $(C',\varg)$, i.~e.\ $M_n$ are flat
cylinders with the vertical cuts. Then the spacetime obtained by
gluing for each $n$ the left bank of the cut in $M_n$ to the right
bank of the cut in $M_{n+1}$ will be just a desired $\zv{M}$. That it
satisfies any \C\ obeyed by $(C,\varg)$ follows from the fact that
$\zv{M}$ is locally isometric to $(C,\varg)$.
\end{teor}
\subsection{Warning!}
From the next subsection on to make the text readable I omit the
letter \C\ and write just `spacetime', `extension', etc.\ instead of
`\C-spacetime', `\C-extension', etc. This definitely is a misuse of
terms, but perhaps not that awful because:
\par\noindent
1. No confusion must arise, since \emph{nowhere} below these words
are used in their `usual' sense;
\par\noindent
2. As has already been mentioned \C\ is not specified. It may, in
particular, be trivial. Which means that all that is below remains
true even if understood `literally', i.~e.\ if this warning is
ignored. The only problem is that what would be proven in such a case
is not the theorem formulated above, but only its weaker version
(obtained from the original one by omitting \C's);
\par\noindent
3. It is quite easy to `restore the real meaning' of any sentence
below. It suffices to add a \C\ to \emph{each}  of the words
`spacetime', `extension', etc;
\par\noindent
4. One need not keep this warning in mind all the time. If something
is valid for \emph{arbitrary} spacetimes, the fact that it is also
valid for \C-spacetimes is absolutely trivial in most cases. The only
exception is the matters of existence and membership. Of course `$A$
is a spacetime' does not necessarily imply `$A$ is a \C-spacetime'. We
shall encounter such not-absolutely-trivial situation only once --- in
proposition~\ref{max} --- and shall take care to show explicitly that
the relevant spacetime is  a \C-spacetime indeed.
\subsection{Outline of the proof}

Consider an extendible spacetime $M$. Let us first try to find an
extension $M_\vartriangle$ (the reason for such notation becomes
evident later) of $M$ such that all causal loops, if they exist in
$M_\vartriangle$, lie in $M$. If we find such an $M_\vartriangle$ and
if, in addition, it is maximal, the theorem would be proven.
\par
Take an \emph{arbitrary} extension $M^\mathrm{ext}$ of $M$ and
consider its open subset $W\equiv M\cup N$, where $N$ is a
\emph{diamond} neighbourhood of a point $p$ in the boundary of $M$.
The precise meaning of the term `diamond' (given in definition~\ref{elem})
is immaterial at the moment. It is important only that a diamond
neighbourhood exists for any point (see proposition~\ref{nasl1}) and
that diamond neighbourhoods are normal
\cite{Neil}. Clearly $W$ is an extension of $M$. Moreover, $N$
(being normal) does not contain any causal loops. Hence new (i.\ e.\ not
confined to $M$) causal loops may exist in $W$ only if there is a
causal curve through a point of $N-M$ such that both its endpoints lie
in $M$ (see the curve $\lambda$ in figure~\ref{fig:outl}).
\karti{h,t,b}{0.2\textwidth}{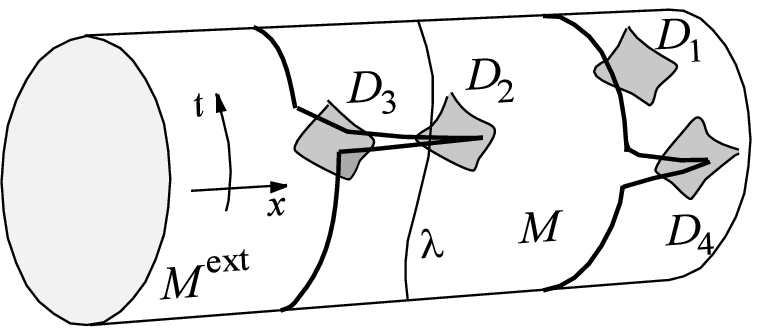}{\label{fig:outl}If we choose
$N=D_3$, $M_N$ will be not causally convex. But we can `unglue' its
lower component from $M$ (so that it is not a part of $M$ any longer
and the dark region in the picture must be viewed as a part of $M$
seen through $N$). $M\cup N$ in such a case have no new causal loops.}
Which means that if we are lucky enough and $M_N\equiv M\cap N$ is
\emph{causally convex in} $N$ (i.~e.\ any causal curve in $N$ lies
in $M_N$ if both its ends do), then we can be assured that $W$ is just
a desired extension $M_\vartriangle$. This condition can be slightly
weakened. Instead of the causal convexity of $M_N$ we can require that
only a connected component $M_\vee$ of $M_N$ is causally convex in
$N$. To obtain $M_\vartriangle$ in such a case one only need `unglue'
all other components of $M_N$ from $M$ (see figure~\ref{fig:outl}).
\par
Generally, $M_\vee$ of course need not be causally convex.
If, for example, $M^\mathrm{ext}$ is  a cylinder $C\equiv\{x\in{}$\R$^1,\:
t=t+1\}$ with the metric $\rmd s^2=\rmd x^2-\rmd t^2$
 and $M$ is the region bounded by the bold line in figure~\ref{fig:outl},
then $M_N$ is causally
convex when $N=D_4$, and is not when $N=D_{1,2}$ (when  $N=D_3$, there are
two connected components in $M_N$, of which the upper one is causally convex
and the lower is not).
 There are spacetimes, however, --- I shall call them
\emph{locally causally convex}, or LCC --- such that $M_\vee$ is causally
convex whatever $M^\mathrm{ext}$ and $M_N$ and whichever component of $M_N$ are
chosen (an example, as can be seen from
proposition~\ref{IGH+sim}, is the Minkowski half-space $t<0$). It follows from the above reasoning that any extendible LCC
spacetime $M$ has an extension $M_\vartriangle$ such that all causal loops
in $M_\vartriangle$ are confined to M (cf.\ proposition~\ref{bound}).
\par
Our next step is constructing in section~\ref{Constr} yet another
extension of an LCC spacetime $M$. This new extension --- denoted by \A\ --- is
made from  $M_\vartriangle$ by
some cutting and gluing and possesses the following properties
(it is the proof in sections~\ref{sec:str},\ref{dvo} of the last two of them
that constitutes the most technical and tiresome part of the whole proof):
\begin{enumerate}
  \item It is locally isometric to $M_\vartriangle$, which in its turn
  is a part of  $M^\mathrm{ext}$. That is how we know that \C\ holds in
  \A\ (which, thus, is a spacetime indeed, see the previous subsection);
  \item Like $M_\vartriangle$ it has no causal loops other than those
  lying in $M$;
  \item Unlike $M_\vartriangle$ it is always locally causally convex
  if such was $M$.
\end{enumerate}
Now that we see that any (extendible) LCC spacetime $M$ can be
extended to a larger (and\emph{ also LCC}) spacetime without the
appearance of new causal loops, the theorem can be proven by just
employing the Zorn lemma as soon as we show that any  $U$ has an LCC
extension $M$ which contains no causal loops outside  $I_M^-(U)$.
This is done as follows. Consider (for a given $U$) the set $\mathsf
V$ of all possible spacetimes of the form $I^-_{U'}(U)$. Clearly, all
causal loops in any $V\in\mathsf V$  lie in the chronology past of $U$
(just because it is the whole spacetime). So, all we need is to find
an LCC element in $\mathsf V$. To this end we show (again by using the
Zorn lemma) that there is a maximal element $V^\mathrm{m}$ in $\mathsf
V$, i.~e.\ such an element that no  $V\in\mathsf V$ is an extension
of $V^\mathrm{m}$. In no extension of $V^\mathrm{m}$ can a past
directed causal curve leave $V^\mathrm{m}$. Whence $V^\mathrm{m}$ is
LCC.
\subsection{Notation and conventions}

In this paper the signature is chosen to be $(-,+,+,+)$.  Whenever
possible I use capital Latin letters of different fonts to denote 4-
and 3-dimensional sets ($U$, $M$, etc.\ for the former and $\EuScript
B$, $\EuScript S$, etc.\ for the latter), and Greek capital letters to
denote 2- and 1-dimensional sets of points. Small Greek and Latin
letters will denote, as a rule, curves (and sometimes isometries) and
points, respectively. Also the following notation will be used:
$$
{<} p, r{>}^{\phantom+}_ U\equiv I^+_U(p) \cap I^-_U(r),\quad
{\leqslant} p, r {\geqslant}^{\phantom+}_U\equiv J^+_U(p) \cap J^-
_U(r),
\quad\lrj{p,r}\equiv \lrj[M]{p,r},
$$
where $U$ is  an open subset of a spacetime $M$.
\begin{teor}{Remark}
\label{nasl} If  $a,b\in  \lr[U]{x,y}$, then
obviously
\[
 \lr[{{\lr[U]{x,y}}}]{a,b}=\lr[U]{a,b}\quad
\ftext{ and }\quad \lrj[{{\lr[U]{x,y}}}]{a,b}=\lrj[U]{a,b}.
\]
\end{teor}

\section{Various types of sets}

In this section I introduce the notions of `diamond' and `locally
causally convex' sets and for later use establish some basic
properties
of such sets. Some of the material of the section
(definitions~\ref{normal},
\ref{cconv}, \ref{simple} and --- most
likely --- propositions \ref{grvyp}, \ref{simsets}) can be found
elsewhere and is included to make the paper self-contained.
\subsection{(Causally) convex sets}

\begin{teor}{Definition}
\label{normal} An open set $O$ is  \emph{convex} if it is a
normal neighbourhood of each of its points.
\end{teor}
With any two points $x,y$  a convex set $O$
contains also a (unique) geodesic segment $\lambda_{xy}$ that
connects
them. To an extent this property is shared by the closure of a convex
set.
\begin{teor}{Proposition}
\label{grvyp} If $O$ is a convex subset of a convex
spacetime $M$, $a\in O$  and $c,d\in \overline{O}$,
then
\[
\lambda_{ac} - c\subset O,\quad
\lambda_{cd}\subset \overline{O}.
\]
\end{teor}

\begin{teor}{Definition}
\label{cconv} An open set $O$ is called a \emph{causally convex
subset} of $U$ if with any two points $a,b$ it contains also the set
$\lr[U]{a,b}$.
\end{teor}
Note that in contrast to convexity, causal convexity is not an
intrinsic property of a set. That is, if $U_1$ is a convex and
causally convex subset of $M$, then any $U_2\subset M$ isometric to
$U_1$ is also convex, but not necessarily causally convex.
\subsection{Simple sets}
\begin{teor}{Definition}
\label{simple} An open set  $O$ is \emph{simple} if it is  convex
and its closure is a compact subset of some other
convex neighbourhood.
\end{teor}
The Whitehead theorem ensures  the existence of  a
simple neighbourhood of any point of any spacetime. Since a
neighbourhood
of a point is  itself a spacetime (and since any its simple subset is
at the same time a simple subset of the larger spacetime), this
means that any point has `arbitrarily small'  simple neighbourhoods,
or
in other words that the simple neighbourhoods constitute a base of
topology
in any spacetime.
\begin{teor}{Proposition}
\label{simsets} Let $O$ be convex. Then any connected component
of $O\cap O'$ is convex (simple), if so does $O'$. Also if $O$ is
simple and its subsets $O_1,O_2$ are convex, then the following sets
\[
O_1,\quad O_2,\quad O_1 \cap O_2
\]
all are simple.
\end{teor}

Simple sets are still `not simple enough' for our needs. The problem
is that there is no direct relation between simplicity and causal
convexity. A timelike curve (provided it is not geodesic) leaving a
simple set still can return in it.
Below we shall overcome this problem by distinguishing a special
subclass of convex sets. In doing so we shall lean upon the following
fact.
\begin{teor}{Proposition}
\label{wit} Any point $q$ of any spacetime has a simple
neighbourhood $O$ such that the sets $I_O^\pm (p)$ are simple for any
$p\in O$.
\end{teor}
\begin{proof}
Let $\{ \bi{e}_{(i)} \}$ with
${e_{(i)}}^\mu{e_{(j)}}_\mu=\eta_{ij}$ be a
smooth frame field in some simple neighbourhood $O'$ of $q$. Emitting
all possible geodesics $\lambda$ from each point of $O'$ we
introduce a normal coordinate system $X^\mu_{\{p\}}$ for each $p\in
O'$ by the following procedure: for any point $r$ we find the
geodesic $\lambda_{pr}(\xi)$, where $\xi$ is an affine parameter such
that $p=\lambda_{pr}(0),\; r=\lambda_{pr}(1)$, and ascribe to $r$
the coordinates $X^\mu_{\{p\}}(r)$ equal to the
coordinates of $\partial_{\xi} (p)$ in the basis $\{ \bi{
e}_{(i)}(p) \}$. The functions $X^\mu_{\{p\}}(r)$ depend smoothly on
both $r$ and $p$. Hence, in particular, for any $\delta$ there
exists a simple neighbourhood $O_\delta$ of $q$  such that
    \begin{equation}
    \label{odelta}  \overline{O_\delta} \subset O', \quad
\sup_{p,r\in
    O_\delta}X^\mu_{\{p\}}(r) < \delta,\ \forall\mu.
    \end{equation}
We choose $O_\delta$ with sufficiently small $\delta$ to be the
desired $O$.
So to prove the proposition we only need to show that
    \begin{equation}
    \label{forml1}
    \exists\, \delta\colon\quad \lambda_{ab} \subset
I_{O_\delta}^\pm (p)
    \qquad \quad \forall p,a,b\colon\quad p\in O_\delta
    ,\quad  a,b\in I_{O_\delta}^\pm (p)
    \end{equation}
(strictly speaking \eqref{forml1} means that $I_{O}^\pm (p)$ are
convex, but proposition~\ref{simsets} ensures that they are simple as
well).
\par
For causal $\lambda$'s \eqref{forml1} follows just from the
definition of $I_{O_\delta}^\pm (p)$, so we can restrict ourselves to
the spacelike ones:
    \begin{equation}
    \label{norm}
    l^\mu l_\mu = 1,\quad \ftext{where {}}
    l_\mu \equiv (\partial_\tau)_\mu.
    \end{equation}
Here $\tau$ is understood to be an  affine parameter.
Let us introduce the function
\[
    \sigma (r) \equiv
    \varg(\bi{x}(r),\bi{x}(r)),
\]
where $\bi{x}(r)\in T_r$ is the `position vector' \cite{Neil}
defined for a fixed $p$ by $x^\mu(r)=X^\mu_{\{p\}}(r)$.  Since
$\sigma$ is a smooth function negative inside $I_{O_\delta}^\pm (p)$
and positive  outside,
\eqref{forml1} will be proved once we prove
that, when $\delta$ is small enough, a spacelike geodesic can
touch a null cone only from outside:
    \begin{eqnarray}
    \label{conv}\sigma''(\tau_0)>0,
    \\
    \ftext{where }\;
    \sigma(\tau)\equiv\sigma\circ\lambda_{ab}\,(\tau),\quad'\equiv
    \frac{\rmd}{\rmd\tau},\quad \tau_0\colon\;
    \sigma(\tau_0)=0,\; \sigma'(\tau_0)=0 .
    \nonumber\end{eqnarray}
To obtain \eqref{conv} let us first use the
relation
$\sigma,_\mu=2x_\mu$ (proved e.~g.\ in \cite{Neil})
    \begin{equation}
    \label{s'} \sigma'=\sigma,_\mu l^\mu=2x_\mu l^\mu.
    \end{equation}
This gives
\begin{equation}
\label{s''} \sigma''= (2x_\nu l^\nu),_\mu l^\mu=
2 l^\nu l^\mu x_{\nu;\mu}
\end{equation}
(in the last equation we used the fact that $\lambda_{ab}(\tau)$ is a
geodesic).
Now consider the point $\lambda_{ab}(\tau_0)$, where $\lambda_{ab}$
touches the null cone, and   for $\bi{z}\in
T_{\lambda_{ab}(\tau_0)}$
denote by $ T^\bot_\bi{z}$ the 3-dimensional subspace of
$T_{\lambda_{ab}(\tau_0)}$ orthogonal to $\bi{z}$. It follows from
(\ref{s'}) that $\bi{l}(\tau_0)\in
T^\bot_\bi{x}$. Also $\bi{x}\in T^\bot_\bi{x}$, since
$\bi{x}$
is null (recall that $\sigma(\tau_0)=0$). Take
a basis $\{\bi{x},\bi{d}_{(1)},\bi{d}_{(2)}\}$ in $
T^\bot_\bi{x}$ defined (non-uniquely, of course) by the following
relations:
    \begin{equation}
    \bi{d}_{(\alpha)} \in T^\bot_\bi{x}\cap T^\bot_{\bi{
e}_{(0)}}
     ,\quad  \varg( \bi{d}_{(\alpha)},\bi{d}_{(\beta)}
)=\delta_{\alpha\beta}
    \qquad \alpha,\beta=1,2. \label{basis}
    \end{equation}
Decomposing $\bi l$ we see from \eqref{norm} that
    \begin{equation}
    \label{l()} { l^{(1)} }^2 + { l^{(2)} }^2 = 1,
    \quad\ftext{where {}}
    \bi{l}=
    l^{(1)}\bi{d}_{(1)} + l^{(2)}\bi{d}_{(2)} +
l^{(3)}\bi{x} .
    \end{equation}
The term $l^{(3)}\bi{x}$ in $\bi l$  gives no
contribution to the right hand side of \eqref{s''} because $x^\nu
x_{\nu;\mu}=\sigma,_\mu/2 =x_\mu$ and $x^\mu x_{\nu;\mu} $ is
proportional to $x_\nu$ (since $\bi{x}$ is tangent to a geodesic).
Hence
    \begin{equation}
    \label{s''2}
    \sigma''=2 l^{(\alpha)} l^{(\beta)} d_{(\alpha)}^\nu
d_{(\beta)}^\mu
    x_{\nu;\mu}
    =2 l^{(\alpha)} l^{(\beta)} d_{(\alpha)}^\nu d_{(\beta)}^\mu
    (g_{\mu\nu} + \Gamma_{\nu,\mu\rho}x^\rho).
    \end{equation}
With the notation
    \[
\Gamma\equiv
2\big| l^{(\alpha)}l^{(\beta)} d_{(\alpha)}^\nu d_{(\beta)}^\mu
\sup_{\nu,\mu,\rho\atop r,p\in
O'}\Gamma_{\nu,\mu\rho}\big|
    \]
we find from (\ref{odelta},\ref{basis},\ref{l()},\ref{s''2}) that
    \[
    |\sigma'' - 2|\leq  4\Gamma \delta.
    \]
$l^{(\alpha)}$ are bounded [see (\ref{l()})] and so
obviously are $d_{(\alpha)}^\nu$.
Thus $\Gamma$
is finite and hence $\sigma''$ is positive for sufficiently small
$\delta$, which proves \eqref{conv} and thereby the whole
proposition.
\end{proof}

\subsection{IGH neighbourhoods}
%
A few useful characteristics of a set are obtained by simply
forgetting
about the ambient space.
\begin{teor}{Definition} Let $U$ be an open subset
of a spacetime $(M,g)$. We call $ U$ \emph{intrinsically (strongly)
causal} if $({U},\ogr{g}{ U})$ is a (strongly) causal spacetime and
\emph{intrinsically globally hyperbolic} (IGH) if $({U},\ogr{g}{ U})$
is
globally hyperbolic.
\end{teor}
\begin{teor}{Proposition} Any point $q$ of any simple set $O$ has an
IGH
neighbourhood $U$ of the form
\label{vggo} $\lr[O]{p,r}$, where $p,r\in O$.
\end{teor}
\begin{proof}
Consider two sequences $p_m,r_m\in O$:
\[
p_m ,r_m \to q ,
\quad  q\in Q_m\equiv{\leqslant} p_m ,r_m {\geqslant}^{\mathstrut}_O.
\]
Each set $ Q_m$ is closed in the topology of $O$ \cite[prop.
4.5.1]{HawEl}, but it well may be not closed in the topology of the
ambient spacetime $M$. When the latter is true there exists a point
$x_m
\in \overline {Q_m}\cap \Bd
O$.  If $x_m$ would exist for infinitely many $m$, there would be a
subsequence $x_n$ converging to some $x\in
\Bd O$ and the sequences of geodesics $\lambda_{p_n x_n}$ and
$\lambda_{r_n x_n}$ (of which the former are future- and the latter
are past-directed) would converge to the same geodesic
$\lambda_{qx}$,
which thus would be both future- and
past-directed at once. This is impossible and hence there exists
$m_0$ such
that $Q_{m_0}$ is closed in the topology of $M$, or in other words
(recall that $\overline{O}$ is compact)
    \[
    \exists m_0:\quad Q_{m_0}={\leqslant}p,r
    {\geqslant}^{\mathstrut}_O\quad\ftext{is compact,
    \qquad where }
     p\equiv p_{m_0},\; r\equiv r_{m_0}.
    \]
Being a subset of the simple neighbourhood $O$ the set
$U\equiv\lr[O]{p,r}$ is intrinsically strongly causal
\cite[prop.~4.10]{Penr}. So, to prove that it is IGH it remains only
to show
that $\lrj[U]{a,b}$ is compact for any $a,b\in U$. Which follows
from the fact (see remark~\ref{nasl}) that
    \[
    \lrj[U]{a,b}=\lrj[O]{a,b}
    \]
and the right hand side is a closed subset of the
compact  $Q_{m_0}$.
\end{proof}

\subsection{Diamond sets}
\begin{teor}{Definition}
\label{elem} Let $\mathsf R(D)$ be a set consisting of a spacetime
$D$ and all its subsets of the form $\lr{x,y}$. We call $D$
\emph{diamond}\footnote[1]{The name refers to the shape of an obvious
diamond spacetime --- the region $\lr{a,b}$, where $a$ and $b$ are
some points in the Minkowski plane. After this paper had
been written I recalled that Yurtsever already used the term
`diamond'
to denote another type of sets \cite{Yur}. The contexts
however are so different that no confusion must arise.} if
any $A\in \mathsf R(D)$:
\par\noindent(\emph{i}) is convex; (\emph{ii}) is IGH; and
(\emph{iii}) with any two points $a,b$ contains also points $c,d$
such that $a,b\in\lr[A]{c,d}$.
\end{teor}

\begin{teor}{Remark}
\label{nasl1}It follows from remark~\ref{nasl} that $A\in \mathsf
R(D)$ implies $\mathsf R(A)\subset \mathsf R(D)$. Therefore, if $D$ is
IGH or diamond then so does any $A\in \mathsf R(D)$.
\end{teor}

With so many good qualities diamond sets are, in the Lorentzian case,
a good candidate for the role fulfilled in the Riemannian case by
balls. The following proposition shows that they also constitute a
base of topology of spacetime.
\begin{teor}{Proposition} Any point $q$ of any spacetime has a
diamond
neighbourhood.
\label{elemexis}
\end{teor}
\begin{proof}
Let $O$ and $U\subset O$ be the neighbourhoods of $q$ from
propositions~\ref{wit} and \ref{vggo}, respectively.
 Then any $A\in \mathsf R(U)$ is,
first, simple (by propositions~\ref{simsets},\ref{wit}) and, second
IGH (by
proposition~\ref{vggo} coupled with remark~\ref{nasl1}).  Condition
(\emph{iii}) obviously also holds in $A$. So $U$ can be taken as the
desired neighbourhood.
\end{proof}

\subsection{Locally causally convex sets}
Now we are in position to introduce the notion that plays the central
part in our proof --- local causal convexity, which is an analog of
causal convexity, but in contrast to the latter characterizes
the set itself and does not depend on the way it is embedded into a
larger space.
Let $D$ be a diamond subset of an extension $M^e$ of a spacetime $M$,
and $D_\vee$ be a connected component of $D \cap M$.
\begin{teor}{Definition}
\label{good}  $M$ is \emph{locally causally convex} (LCC) if for
any $M^e$ and $D$ each $D_\vee$ is a causally convex subset of $D$.
\end{teor}
Generally, neither convex, nor IGH sets are LCC (consider a rectangle
in the Minkowski plane, and the `bad' set from \cite{loc-n-cau},
respectively). However the following holds.

\begin{teor}{Proposition}
\label{IGH+sim} Any IGH convex spacetime $M$ is LCC.
\end{teor}
\begin{proof}
Consider a future-directed timelike curve $\gamma\subset D$ from $a$
to $b$, where $a,b \in D_\vee$. We must show that the whole $\gamma$
lies in $D_\vee$.
\par
Let $c\in \gamma \cap D_\vee$.
By proposition~\ref{simsets} $D_\vee$ is convex. Therefore a geodesic
$\lambda_{ac} \subset D_\vee$ from $a$ to $c$ must exist.
$\lambda_{ac}$ is future-directed [$D$ is convex,  so
$\lambda_{ac}$ is the unique geodesic connecting $a$
and $c$ in $D$, while the existence of $\gamma$ guarantees that $c\in
I_D^+(a)$]
and thus $c\in \gamma \cap J_{ D_\vee}^+(a)$. Similarly, $c\in \gamma
\cap J_{ D_\vee}^-(b)$, and so,
\[
\gamma \cap D_\vee =  \gamma \cap \lrj[D_\vee]{a,b}.
\]
The right hand side is compact (since both $M$ and $D$ are IGH and
convex), and $ D_\vee$ is open. Consequently, $\gamma \subset
D_\vee$.
\end{proof}
\begin{teor}{Corollary}\label{cor:diamlc}
Any diamond spacetime is LCC (and, accordingly, any point has an
arbitrarily small intrinsically causal LCC neighbourhood).
\end{teor}
The reverse, of course, is not true. One of the reasons is that local
causal convexity is, loosely speaking, a characteristic of the
`superficial' (i.~e.\ lying `near the boundary') regions of a
spacetime rather than its bulk. This, in particular, entails quite
regular structure of the boundary of an LCC region:  `mostly' it is
achronal (though not \emph{always} by the reasons obvious from
inspection of  $D_4$ in figure~\ref{fig:outl}).
\begin{teor}{Proposition}
\label{achrgr} If $M$ is an LCC subset of a spacetime $M_1$ and $U$
is a neighbourhood of some point of $\Bd M$, then within $U$ there
always exists a diamond set $H$ such that, $\Bd \Hl\cap H$ (where
$\Hl$ is a connected component of $H\cap M$) is a closed, imbedded,
achronal three-dimensional $C^{1-}$ submanifold in $H$.
\end{teor}
\begin{proof}
We begin by proving that there is a timelike curve in $M$ whose end
point lies in $\Bd_{U} M$  (the boundary of $M$ in $U$).
 Let $U_M$ be  a connected component of $ U\cap M$ and let $p\in
\Bd_{U} U_M$ (see figure~\ref{treug}).
\karti{b,t}{0.5\textwidth}{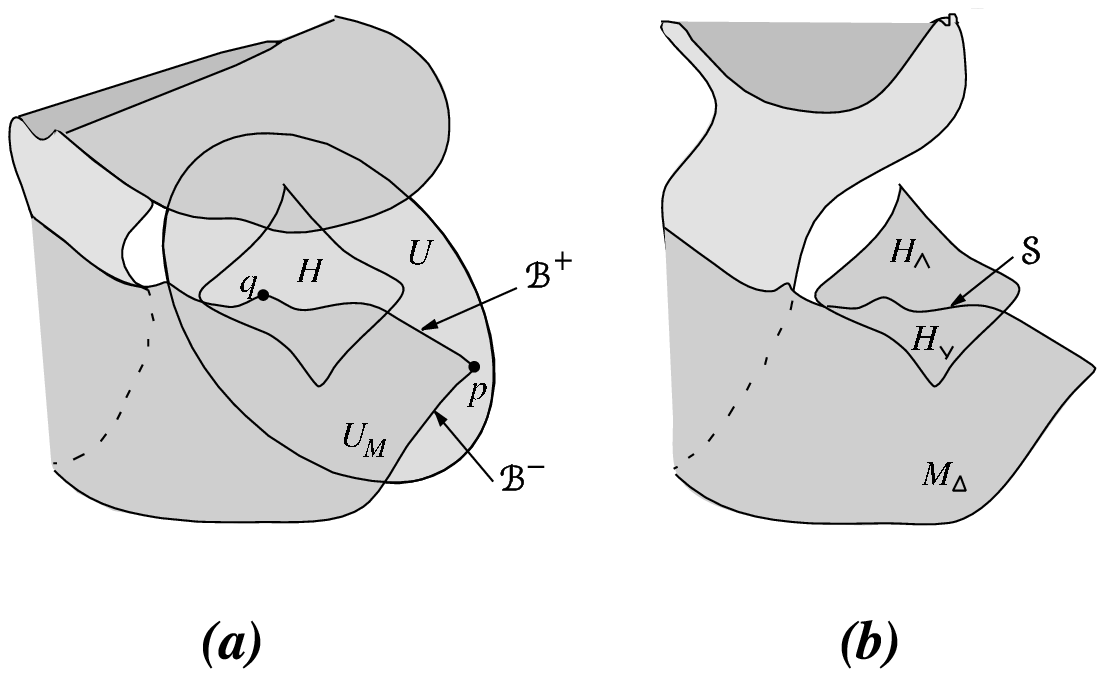}{\label{treug}Of the two  curves
meeting in $p$ the upper one is $\EuScript B^+$ and the lower one is
$\EuScript B^-$. $p$ does not belong to either.}
Denote by $\EuScript B^\pm$ the (maybe empty) sets of points of
$\Bd_{U} U_M$
which are the future (past) endpoints of timelike curves lying in
$U_M$.

Clearly, any timelike curve in $U$ connecting a point $a\in U_M$ with
$p$ contains at least one point of $\EuScript
B^+\cup\EuScript B^-$.  In particular, if $\EuScript
B^+=\varnothing$, then
$I^-_{U}(p)\subset U - U_M$. And hence if
also $\EuScript B^-=\varnothing$, then
$I^+_{U}\big(I^-_{U}(p)\big)\subset U - U_M$, which is impossible
since $I^+_{U}\big(I^-_{U}(p)\big)$ is a neighbourhood of $p$. So
\[
\EuScript B^- \cup \EuScript B^+\neq \varnothing.
\]
Suppose for definiteness that it is $\EuScript B^+$ that
is non-empty:
    \[
    \EuScript B^+\neq \varnothing.
    \]
Our next step is to show that $\EuScript B^+$ and $\EuScript B^-$ are
separated. Let $q\in \EuScript B^+$. By the definition of $\EuScript
B^+$, $q$ lies in $I^+_{U}(U_M)$ and hence so does some its
neighbourhood $Q$. Without loss of generality (see
proposition~\ref{elemexis}) $U$ can be taken
diamond. Then $q\notin \overline{\EuScript B^-}$ since otherwise some
points of $\EuScript B^-$ also would lie in $Q$. Such points would
belong to $X\equiv\big(I^+_{U}(U_M)
\cap I^-_{U}(U_M)\big) - U_M$, but $M$ is LCC, $U$ is diamond, and
so, $X$ must be  empty by definition~\ref{good}. Thus
    \begin{equation}
    \label{empty}
    \EuScript B^\pm \cap \overline{\EuScript B^\mp} = \varnothing.
    \end{equation}
It follows that for a  sufficiently small neighbourhood
$H\subset U$ of $q$
    \begin{equation}
    \label{h} H\cap \EuScript B^- =\varnothing.
    \end{equation}
To complete the proof it suffices now to require that $H$ be diamond
(which by proposition~\ref{elemexis} is always possible) and to
choose as $\Hl$ a component of $H\cap M$ lying in $U_M$. Any
past directed timelike
curve in $H$ leaving $\Hl$  (and thus also $U_M$) would
contain a point of $\EuScript B^-$, which is impossible by \eqref{h}.
So $\Hl$ is the past set in $H$ and hence by
\cite[proposition~6.3.1]{HawEl}
its boundary is a closed, imbedded, achronal three-dimensional
$C^{1-}$ submanifold of $H$.
\end{proof}
%
%
\section{Construction of \A}

In this section \label{Constr} for an arbitrary extendible LCC
spacetime $M$
 we construct an extension $\A$ of a
special type (as will be proved in the subsequent sections $\A$ is
LCC and
has no closed causal curves except those lying in $M$). $\A$ will be
built in
a few steps. First we glue a
diamond region $H$ to $M$ obtaining thus an extension
$M_\vartriangle$ (see figure~\ref{treug}b). Then to the `upper' (that
is lying outside $M$) part of $M_\vartriangle$ we glue yet another
copy of $H$ (in doing so we  remove a three-dimensional surface,
so that the resulting spacetime $\M$ (depicted in
figure~\ref{mdiam}a)
be Hausdorff). Finally, a smaller diamond set $H'$ is glued to $\M$
(see figure~\ref{mn}).
\subsection{The spacetime $M_\vartriangle$.}

\begin{teor}{Proposition} Any extendible LCC spacetime $M$ has an
extension $M_\vartriangle$ such that: \par\smallskip\noindent
(I). $M_\vartriangle=M\cup H$, where $H$ is diamond, and $M\cap H$ is
connected;
\par\noindent
(II). $M$ is a past set in   $M_\vartriangle$, or (see
remark~\ref{II} below)
\par\noindent
(II$'$). $M$ is a future set in   $M_\vartriangle$.
\label{bound}
\end{teor}
\begin{proof}
Let $M_1$, $H$, and  $\Hl$ be as in proposition~\ref{achrgr}. Let further
$\overset{*} M$ and $\overset{*} H$ be spaces isometric to $M$ and
$H$, respectively:
\[
\phi_M\colon\;M \to \overset{\: *} M, \qquad \phi_H\colon\; H
\to\overset{\: *} H,
\qquad\phi_M,\phi_H\ftext{{} --- isometries.}
\]
The spacetime $M_2 \subset M_1$ defined by $M_2\equiv M\cup
H$ can be presented as a result of `gluing' $\overset{\: *} H$ to
$\overset{\: *}
M$ by an isometry:
\[
M_2 = \overset{\: *} M \cup_\phi \overset{\: *} H,
\]
where
    \begin{equation*}
     \phi\equiv {\phi_M}\circ{\phi_H }^{-1}\colon\quad
     {\phi_H}(H \cap M)  \to  {\phi_M}(H \cap M).
    \end{equation*}
We construct $M_\vartriangle$ by `ungluing' $\overset{\: *} H$
from $\overset{\: *} M$ along all but $\Hl$ connected components
of $H \cap M$ (see figure~\ref{treug}):
\[
M_\vartriangle\equiv \overset{\: *} M \cup_{ \ogr{\phi}{ \phi_H(\Hl)
} } \overset{\: *} H .
\]
From now on by $M$, $H$, and $\Hl$ we understand the corresponding
regions of $M_\vartriangle$ (this must not lead to any confusion,
since we shall not consider $M_1$ any more). $M_\vartriangle$ is
locally isometric to a part of $M_1$ and hence \C\
holds in it. So, it is a \C-extension of $M$, indeed. Further, $M_1$
satisfies (I) by construction and (II), or (II$'$) (depending
on whether $q$ was taken in  $\EuScript B^+$, or in $\EuScript B^-$)
by the reasons discussed in proposition~\ref{achrgr}.
\end{proof}
\karti{t}{0.40\textwidth}{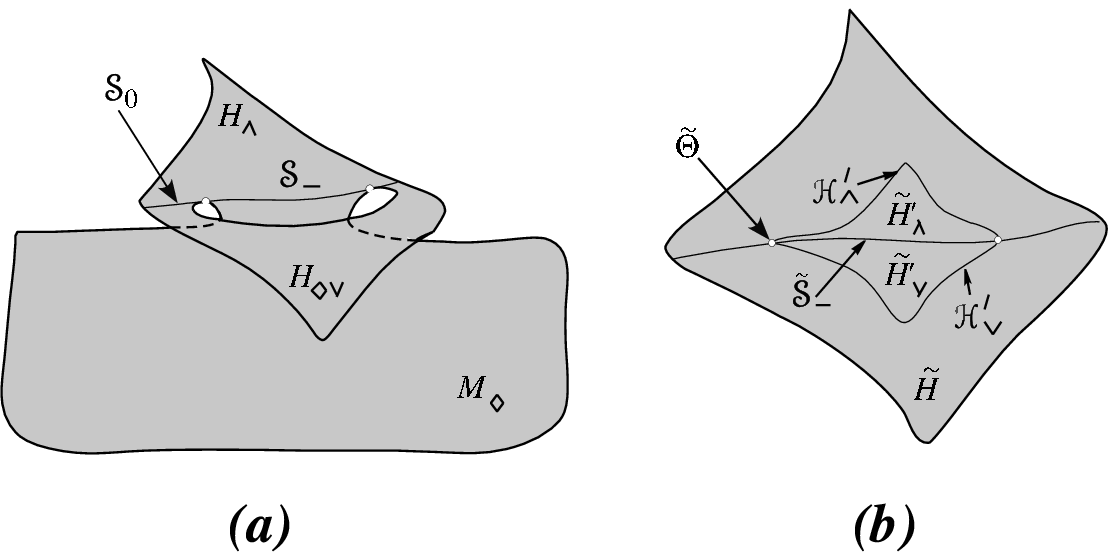}{\label{mdiam}$\M$ is obtained by
gluing $\tilde K=\tilde H-\overline{\tilde\EuScript S_-}$ to
$M_\vartriangle -\overline{ \EuScript
S_0}$.}

\begin{teor}{Remark}
In what follows we assume for definiteness that it is (II)
that holds for our $M_\vartriangle$. \label{II}
\end{teor}

\begin{teor}{Remark}
Condition (II) of course implies that the boundary $\EuScript S\equiv
\Bd M$ is a closed, imbedded, achronal
three-dimensional $C^{1-}$ submanifold in $M_\vartriangle$. This
surface divides $H$ into two parts: $H_{\vee}$ and $H_{\wedge}\equiv
H - \overline H_{\vee}$.
\end{teor}

\begin{teor}{Remark}
\label{I} By the definition of local causal convexity condition (I)
implies that all causal loops in $M_\vartriangle$ (if there are any)
are confined to $M$. If in addition $M_\vartriangle$ always were LCC,
which unfortunately is not the case, we would not need anything below
up to proposition~\ref{max}.
\end{teor}

\subsection{The spacetime $\M$.}

Now we want to construct for $M$ yet another extension, which we shall
denote by $\M$ (and which is not an extension of $M_\vartriangle$).
We shall do this similarly to the way we constructed $ M_\vartriangle$
that
is by, first, presenting some auxiliary spacetime ($
M_\vartriangle - \Theta$, see below) as a result
of gluing  together two spacetimes ($\tilde K$ and $\hat
M_{{\bowtie}}$)
and by  then ungluing them along a connected component of their
intersection.
\par
Let $H'\subset H$ be a diamond neighbourhood with the compact closure
in
$H$ and let it intersect $ \EuScript S$, thus splitting the latter
into
three non-empty disjoint parts (see figure~\ref{mdiam}):
\[
\EuScript S_-\equiv\EuScript S\cap H',\qquad
\EuScript S_0 \equiv \EuScript S - \overline{\EuScript S_-},\qquad
\Theta\equiv \overline{ \EuScript S_-} - \EuScript  S_-.
\]
$\EuScript S_-$ in its turn divides $H'$ into  two disjoint regions:
\[\Ol\equiv H'\cap\Hl,\qquad  \Ou\equiv H'\cap\Hu.
\]

Now let $\hat M_\vartriangle$ and $\tilde H$ be spaces isometric to
$M_\vartriangle$ and $H$, respectively. The isometries are:
\[
\psi_M\colon\;M_\vartriangle \to {\hat M_\vartriangle},
 \qquad \psi_H\colon\;H \to\tilde  H.
\]
We shall often write $\tilde A$ for $\psi_H(A)$ and sometimes $\hat
A$
for $\psi_M(A)$. In particular:
\[
\tilde \Hu\equiv  \psi_H(\Hu),\quad\Olt\equiv \psi_H(\Ol ),
\quad\tilde \EuScript S_-\equiv \psi_H( \EuScript S_- )
,\quad\hat \EuScript S_- \equiv\psi_M( \EuScript S_- ),\ftext{{}
etc.}
\]
Consider the spacetime $M_\vartriangle - \Theta$. Obviously,
\[
M_\vartriangle - \Theta = \hat M_{{\bowtie}} \cup_{\psi_{MH}} \tilde
K,
\]
where
\[
    \qquad M_{{\bowtie}} \equiv M_\vartriangle - \overline{
\EuScript
S_0},
    \qquad  K\equiv  H - \overline{\EuScript S_-} , \qquad
\psi_{MH} \equiv \psi_M\circ{\psi_H}^{-1}.
\]
Note that $M_{{\bowtie}}\cap K$ consists of two disjoint regions:
$\Hl$ and $ \Hu$, which enables us to build a new spacetime by
ungluing $K$ from $M_{{\bowtie}}$ along one of them. Namely, we
define
    \begin{equation}
    \M \equiv \hat M_{{\bowtie}} \cup_\psi \tilde K,\qquad
\ftext{where }
    \psi \equiv \ogr{\psi_{MH}}{\tilde \Hu}.
    \label{gl}
    \end{equation}
The definition \eqref{gl} produces two natural
isometries:
\[
\varpi_{M_{{\bowtie}}}\colon\:  \hat M_{{\bowtie}} \to \M,\qquad
\varpi_K\colon\:  \tilde K\to \M.
\]
From now on for the subsets $ \varpi_{M_{{\bowtie}}}( \hat
M_{{\bowtie}})$, $\varpi_{M_{{\bowtie}}}( \hat K)$,
$\varpi_{M_{{\bowtie}}}( \hat \EuScript S_-)$, etc.\ of $\M$ we shall
write\footnote{Strictly speaking, this is some abuse
of notation because originally
we took, say, $ M_{{\bowtie}}$ to be a part of $M_{\vartriangle}$
(not of $\M$) but
it must not lead to any confusion, since we shall not consider
$M_{\vartriangle}$ any more.} simply $ M_{{\bowtie}}$, $K$,
$\EuScript S_-$, etc., while the images of $\varpi_K$ we
shall mark by $\lozenge$, i.~e.\ $\varpi_K(\tilde K)$,
$\varpi_K(\tilde H'_\vee)$, $\varpi_K(\tilde H_\vee)$, etc.\ we shall
denote by $\K$, $H'_{\lozenge\vee}$, $H_{\lozenge\vee}$, etc.
In this notation $ M_{{\bowtie}}\cap\K=\Hu$, for example.

\subsection{The spacetime \A.}
We shall be interested in one particular type of extensions of $\M$,
which we obtain   by pasting one more copy of $H'$ to $\M$. More
specifically, we take a spacetime $ H'_{\mathcal N}=\varsigma (\tilde
H')$ (where $\varsigma$ is an isometry) and define a new spacetime as
follows
\[
\A\equiv\M\cup_\chi H'_{\mathcal N},\qquad\ftext{where }
\chi\equiv\ogr{\varsigma\circ{\varpi_K}^{-1}}{H'_{\lozenge\vee}}.
\]
\karti{t}{0.40\textwidth}{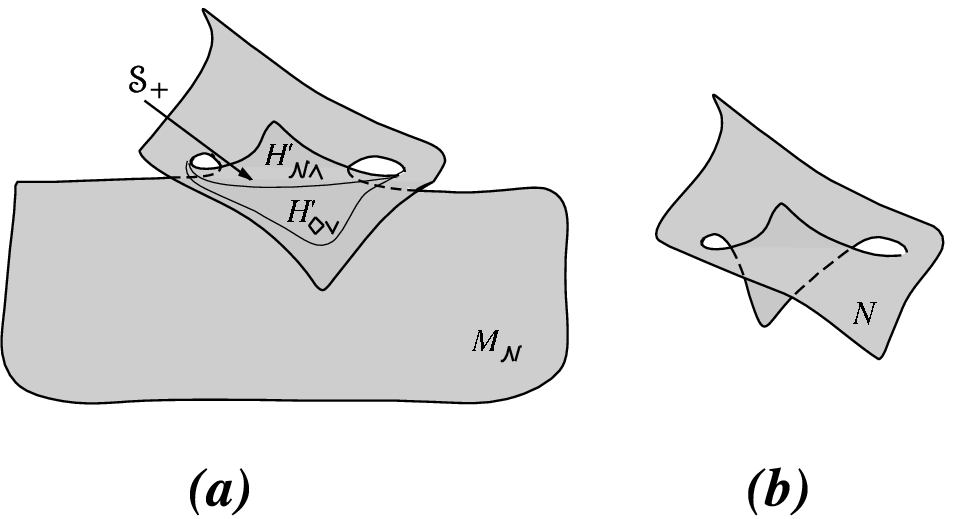}{\label{mn} $\A$ can
be viewed as a result of gluing $N$ to $M$.}
 It is easy to see that $\A$ is an extension of $\M$
[indeed, $
\A- \overline{\M}=\Ola\equiv\varsigma (\tilde H'_\wedge)$, so it is an extension
of $\M$ in the `usual' sense (see the introduction) and
 \C\ holds in $\A$ since the latter is locally isometric to the
corresponding part of $M_1$, cf.\ proposition~\ref{bound}].

\begin{teor}{Remark}
\label{nakr}
Consider the region $N\subset \A$ (depicted in figure~\ref{mn}) and
the projection $\pi\colon\:N\to\tilde H$ defined as
follows:
\[
N \equiv \A - (M - \Ol)=\overline{\Ol}\cup\K\cup\overline{\Ola},
\]
\[
\ogr{\pi}{H'}\equiv ( \varpi_{ M_{\bowtie}
}\circ\psi_{MH} )^{-1},\quad
\ogr{\pi}{\K}\equiv{\varpi_K}^{-1},\quad
\ogr{\pi}{H'_{\mathcal N}}\equiv {\varsigma}^{-1}.
\]
Clearly $N$  is an area in the universal covering
of
$\tilde H - \tilde\Theta$.
\end{teor}

\section{The structure of \A}\label{sec:str}
Our proof in section~\ref{dvo} of that $\A$ is LCC will be based on
the fact that in its extensions some curves with the same ends
are nonhomotopic, and thus cannot belong to the same diamond set.
What makes the curves nonhomotopic is a singularity that is
present in $\A$ in spite of the fact that $\A$ was assembled
of a few spacetimes each free from singularities. The nature of
this singularity is precisely the same as in the Deutsch-Politzer
spacetime, or, say, in the double covering of the Minkowski
plane with a deleted point.

In this section we,
first, establish that indeed there is a singularity in $\A$ and then
consider, among other things, the ensuing restrictions on homotopic
curves. Some of the facts concerning intersection of curves with the
surfaces $\EuScript S_\pm$ and $\Blu$ seem obvious, but have to be
\emph{proved} because these surfaces are not sufficiently smooth
to fall under the standard results.

\subsection{The singularity}

Let us present $\A$ in the following form:
\begin{equation}
\label{A}
\A =M_{\bowtie}\cup\K\cup H'_{\mathcal N}.
\end{equation}
By construction (see proposition~\ref{bound}) $M$ is a past set in
$M_{\bowtie}$. So (since
$\K\cup H'_{\mathcal N}$ is open and has no common points with
$M$) $M$ is a past set in $\A$ as well. Similarly, $\Ola$ is a future
set in $ H'_{\mathcal N}$ and has no common points with
$M_{\bowtie}\cup\K$ and hence is a future set in $\A$ (and likewise
it
can be shown that $\Hu$ and $H_{\lozenge\vee}$ are, respectively, a
future and a past sets in \A). Which means
that both $\EuScript S_-=\Bd M$ and $\EuScript S_+\equiv\Bd \Ola$ are
imbedded three-dimensional $C^{1-}$ submanifolds achronal in $\A$.
The intersection of $\EuScript S_+$ and $\EuScript S_-$ is
empty,
their union we shall denote by $\EuScript S_{\Bumpeq}$. The following
proposition, loosely speaking, says that $\Theta$ cannot be glued back
into the spacetime.

\begin{teor}{Proposition} In any extension $M^e$ of the spacetime
$\A$ the sets $\EuScript S_\pm$ are closed.
\end{teor}
\begin{proof}
It is easy to
check
that $\EuScript S_+$ is closed if so does $\EuScript S_-$. So we
shall only prove the proposition for the latter. Suppose
$\{a^{(1)}_n\}$ is a sequence of points such that (contrary to our
claim):
\[
a^{(1)}_n\in \EuScript S_-,\qquad \exists\lim a^{(1)}_n=a\notin
\EuScript S_-
\]
and $\hat a^{(1)}_n$ are the images of these points in $\hat
M_\vartriangle$:
\[
\hat a^{(1)}_n\equiv \psi_M ( a_n^{(1)}).
\]
$\hat a^{(1)}_n$ are confined to the compact set
$\overline{\psi_M(\EuScript S_-})$ and therefore there exist points
$x$, $\hat q $, and $ q $:
\[
x\equiv \lim  \hat a^{(1)}_n ,\quad
\hat q \in I^+_ {\hat M_{\vartriangle}}(x),\quad
q\equiv \psi_M^{-1}(\hat q)\in \A.
\]
Obviously,
$x\in  \psi_M(\Theta)$ and so we can find a sequence $\{ \hat
a^{(2)}_n\}$:
\[
\hat a^{(2)}_n \in \psi_M(\EuScript S_0),\qquad \lim  \hat a^{(2)}_n
= x.
\]
Now let $\{\hat v^{(i)}_n \}$ and $\{ v^{(i)}_n \}$ with $i=1,2$ be
the sequences in $T_{\hat q}$ and $T_{q}$, respectively, defined
by
\[
\hat v^{(i)}_n\equiv\exp^{-1}_{\hat q } (\hat a^{(i)}_n)
,\quad v^{(i)}_n\equiv \rmd{\psi_M}^{-1}[\hat v^{(i)}_n].
\]
Clearly,
\[
\lim v^{(i)}_n=v,
\qquad\ftext{where }v\equiv \rmd{\psi_M}^{-1}[\hat v],
\quad\hat v\equiv \lim \hat v^{(i)}_n.
\]
It follows from
\[
a=\lim[\exp_{q } (v^{(1)}_n)]=
\exp_{q } \{ \rmd{\psi_M}^{-1}\big[
\lim \rmd{\psi_M}[v^{(1)}_n]\big]\} =\exp_q(v)
\]
that the existence of $a$ implies the existence (for sufficiently
small positive $\epsilon$) of a point $a'$:
\[
a'\equiv \exp_q [(1+\epsilon)v]= \lim \exp_{q }
[(1+\epsilon)v^{(2)}_n].
\]
Each point
\[
\exp_{q } [(1+\epsilon)v^{(2)}_n]={\psi_M}^{-1}
(\exp_{\hat q} [(1+\epsilon)\hat v^{(2)}_n])
\]
lies in $M$ (recall that $\hat v^{(2)}_n$ is timelike and $\hat M$ is
a past set in $\hat M_{\vartriangle}$) and hence $a'\in
\overline{M}$. Moreover, since $v$ also is timelike
\[
a'\in M.
\]
But repeating the same reasoning for $\psi_{H}$ and $\tilde
a^{(i)}_n\equiv
\psi_H\circ{\psi_M}^{-1}(\hat a^{(i)}_n)$
instead of $\psi_{M}$ and $\hat a^{(i)}_n$, respectively,
one finds that $a' \in K_{\lozenge}$, which is impossible, because by
construction $K_{\lozenge}$  and $M$ are disjoint.
\end{proof}

%
%
\subsection{Curves intersecting $\EuScript S_{\Bumpeq}$}
\begin{teor}{Definition}
Let $\lambda\colon\;[0,1]\to M^e$ be a timelike curve in an extension
$M^e$ of the spacetime $\A$. We call
$\tau_i$ a \emph{positive (negative) root} if
$\lambda(\tau_i)\in \EuScript S_\pm$. The number of the positive
(negative) roots of $\lambda$ we denote by $n_\pm[\lambda]$.
\end{teor}
Obviously for any future-directed $\lambda\subset\A$ the following
holds:
\begin{subequations}
    \label{zven}
\begin{eqnarray}
    \lambda(0)\in \overline{M},\:\lambda(1)\notin M\quad
    &\Rightarrow &\quad n_-[\lambda]=1,
    \quad\ftext{otherwise }n_-[\lambda]=0
\\
    \lambda(0)\notin \Ola,\:\lambda(1)\in \overline{\Ola}\quad
    &\Rightarrow& \quad n_+[\lambda]=1,
    \quad\ftext{otherwise }n_+[\lambda]=0
\end{eqnarray}
\end{subequations}
 \par
Consider a homotopy $\lambda_\xi(\tau)$ with $\xi\in [0,1]$ such that
$\lambda_\xi$ are timelike and future-directed.  We shall denote the
curves $\ogr{\lambda_\xi}{\tau=\tau_0}$ by $\mu_{\tau_0}(\xi)$.
\begin{teor}{Proposition} If the curves $\mu_{0,1}$ do not intersect
$\EuScript S_\Bumpeq$, then $n_\pm[\lambda_0]=n_\pm[\lambda_1]$.
\label{indec}
\end{teor}
\begin{proof}
Suppose $\tau_0$ is not a root of $\lambda_{\xi_0}$. Then the point
$\lambda_{\xi_0}(\tau_0)$ does not belong to the closed set
$\EuScript
S_\Bumpeq$ and therefore some its neighbourhood $U$ also does not
intersect $\EuScript S_\Bumpeq$. So, around the
point $(\tau_0,\xi_0)$ there exists a rectangle
\[
\square_{\delta\epsilon}\equiv\{\tau, \xi\colon\;
|\tau-\tau_0|<\delta,\,|\xi-\xi_0|<\epsilon\}
\]
that does not contain roots.
\par
Now suppose $\tau_0\neq 0,1$ \emph{is} a root (positive for
definiteness) of
$\lambda_{\xi_0}$.  The surface $ \EuScript S_+$ lies in the
spacetime $\A$,
where it bounds the future set $\Ola$, hence for any (sufficiently
small) $\delta$
\[
\lambda_{\xi_0}(\tau<\tau_0) \in \A-\overline{\Ola}
, \qquad\lambda_{\xi_0}(\tau>\tau_0) \in \Ola,
\]
when $|\tau-\tau_0|<\delta$.

So, when $\epsilon_a$ and $\delta_a$ are sufficiently small, any
segment
$\lambda_{\xi}(|\tau-\tau_0|<\delta_a)$ with $|\xi-\xi_0|<\epsilon_a$
will also lie in $\A$ and have its ends one in  $\Ola$
and the other outside. Which by \eqref{zven} implies that
 in $\square_{\delta_a\epsilon_a}$ there is exactly
one root for each $\xi$.
\par
Thus, any point $a=(\tau_0,\xi_0)$ lies in the center of a rectangle
such
that the number of the roots of $\lambda_{\xi}$ located between
$\tau_0-\delta_a$ and $\tau_0+\delta_a$ does not change for $\xi$
varying between $\xi_0-\epsilon_a$ and $\xi_0+\epsilon_a$. Hence (due
to the compactness of $\lambda_\xi$) such a positive constant
$\epsilon_{\xi_0}$ can be found for any ${\xi_0}$ that
$n_\pm[\lambda_{\xi}]$ does not change for ${\xi}\in
({\xi_0}-\epsilon_{\xi_0},{\xi_0}+\epsilon_{\xi_0})$. As ${\xi}$
varies over a compact set $[0,1]$, its whole range can be covered by a
finite number of such intervals with constant $n_\pm[\lambda]$.
\end{proof}

\begin{teor}{Corollary}
\label{traj} If $\tau_n=\tau_n(\xi)$ is the $n$th root of
$\lambda_{\xi}$,
then $\mu_{\tau_n}(\xi)$ is a continuous curve. $\mu_{\tau_n}(\xi)$
and $\mu_{\tau_k}(\xi)$ are disjoint, when $n\neq k$
\end{teor}

\begin{teor}{Remark}
\label{cor4}
All timelike curves lying in a convex
neighbourhood and connecting the same two points have equal $n_\pm$.
Thus, for a convex neighbourhood we can speak about $n_\pm[pq]$
understanding by it $n_\pm[\lambda_{pq}], $ where $\lambda_{pq}$ is
an arbitrary timelike curve lying in this neighbourhood and
connecting the points $p$ and $q$. We also set $n_\pm[pq]=0$ for
$p=q$.
\end{teor}

\begin{teor}{Remark}
If points $a, b, q_1, q_2$ lie in a convex spacetime, $a,
b\notin\EuScript S_{\Bumpeq}$, and $a\succeq b\succeq q_1,q_2$ (by
$x\succeq y$ we mean $x\succ y$, or $x=y$), then
    \begin{equation}
    n_\pm[{aq_2}] - n_\pm[aq_1] = n_\pm[bq_2]- n_\pm[bq_1] .
    \label{lom}
    \end{equation}
    \end{teor}
%
%

Let $D\subset M^e$ be a diamond set and $D_M$ be a connected
component of $D\cap \A$. The following two propositions will help
us to use the fact that $M$ and $\Ola$ are LCC in reducing the proof
of the local causal convexity of $\A$ to examining ${\K}$.

\begin{teor}{Proposition} If a future-directed
timelike curve $\lambda\subset D$ starting and ending in $D_M$ does
not intersect $\EuScript S_\Bumpeq$,
then its ends lie either both in $M$, or both in $\Ola$, or both in
${\K}$.
\label{kriv}
\end{teor}
\begin{proof}
Consider a curve $\varphi\colon\; [0,1]\to D_M$ connecting the ends
of $\lambda$. Cover $\varphi$ by a finite number of diamond subsets
$F_n$ of  $D_M$:
\[
\varphi \subset \bigcup_{n=1,\dots,N} F_n \subset D_M,\qquad
\varphi(0)=\lambda(0)\in F_1,\quad \varphi(1)=\lambda(1)\in F_N
\]
and pick  $N+1$ points $q_n\notin \EuScript S_\Bumpeq$ such that:
\[
q_1\equiv \lambda(0), \quad
q_{N+1}\equiv \lambda(1) , \qquad q_n \in F_{n-1}\cap F_n
\quad\ftext{ for } n\neq 1,N+1.
\]
For any pair $q_n, q_{n+1}$ (since they belong to the same diamond
set $F_n$) we can find a point $b_n\in I_{D_M}^+(q_n)\cap
I_{D_M}^+(q_{n+1})-\EuScript S_\Bumpeq$.
\par
Pick a point $a$ such that
\[
a \in I^+_{D}(b_n)-\EuScript S_\Bumpeq\qquad \forall n
\]
(it is always possible because $D$ is diamond). Applying
twice
 equation \eqref{lom} we get:
\begin{eqnarray}
\fl 0=n_\pm[\lambda]=n_\pm[a,\lambda(0)] - n_\pm[a,\lambda(1)]
\nonumber\\
=n_\pm[q_1,b_1] - n_\pm[b_1,q_2] +n_\pm[q_2,b_2]-  \dots  -
n_\pm[b_N,q_{N+1}].
\end{eqnarray}
Combining this with  \eqref{zven} we  prove the proposition.
\end{proof}
\begin{teor}{Proposition}
\label{nointe}
The sets $D_M\cap M$, $D_M\cap\K$ and $D_M\cap\Ola$  are connected.
\end{teor}
Before proving the proposition we have to establish a lemma. Let
$\lambda_{\xi}(\tau)$ be a
homotopy considered in proposition~\ref{indec} with an additional
requirement that $\lambda_{\xi}$ for each $\xi$ is a geodesic in $D$.
We denote by $\varphi(\xi)\equiv\lambda_{\xi}(\tau'(\xi))$ a curve
(lying in the surface $\lambda(\xi,\tau)$) defined by a continuous
function $0\leqslant\tau'(\xi)\leqslant 1$ and consider the segment
$\lz_{\xi}$ of $\lambda_{\xi}$ between $\varphi$ and an intersection
of $\lambda_{\xi}$ with $\EuScript S_\Bumpeq$.
\begin{teor}{Lemma}
If  $\varphi(\xi)\subset\A$ and \label{svyaz} the condition
\begin{equation}
\label{vyp}
\lz_{\xi}\equiv\lambda_{\xi}([\tau_n, \tau'])\subset \A
\end{equation}
holds for $\xi=0$ then it holds for all the rest $\xi$ as well.
\end{teor}
\begin{proof}
We shall prove the lemma for  a negative root
$\tau_n$ (the case of a positive root can be handled in much the same
way).
Then \eqref{vyp} is equivalent
(recall that $\Hu$ is a future set and $M$ is a past set in $\A$) to
\begin{equation}
\label{bud}
\lz_{\xi}\subset U\equiv
I^+_{\Hu}(\EuScript S_-)\cup \EuScript S_-\cup M.
\end{equation}
Let $\Xi\subset [0,1]$ be the set of all $\xi$ for which
\eqref{bud} [and hence \eqref{vyp}] holds and let $\xi_*$ be a limit
point of this set:
\[
\xi_*=\lim \xi_k,\qquad \xi_k\in \Xi.
\]

We want to show that $\lz_{\xi_*}\subset U$. For $\tau_n(\xi_*)=
\tau'(\xi_*)$ it is trivial. For $\tau_n(\xi_*) > \tau'(\xi_*)$
(i.~e.\ a past-directed $\lz_{\xi_*}$) it follows from the fact
that both $\lambda_{\xi_*}(\tau_n-0)$ and $\lambda_{\xi_*}(\tau')$
lie in a connected component of the intersection of LCC $M$ and
diamond $D$.  Finally, consider the case $\tau_n(\xi_*) <
\tau'(\xi_*)$ (a future-directed $\lz_{\xi_*}$).
By assumption $\lambda_{\xi_*}(\tau')=\varphi(\xi_*)\in\A$. On the
other hand, it is a limit point of $\lambda_{\xi_k}(\tau')$, so it
lies in $\overline{I^+_{\Hu}(\EuScript S_-)}\cap\A$ and thus in $N$.
The points
$\pi[\lambda_{\xi_*}(\tau_n)]$ and $\pi[\lambda_{\xi_*}(\tau')]$
can be connected in $\tilde H$ by a geodesic  $\gamma$. Since
$\tilde H$ and $D$ are convex
\[
\gamma=\lim\pi(\lz_{\xi_k})=\pi(\lim\lz_{\xi_k})=\pi(\lz_{\xi_*})
\]
and therefore (recall that $\lz_{\xi_*}$ is  timelike)
\[
\gamma - \gamma(\tau_n) \subset I^+_{\tilde\Hu}(\tilde\EuScript S_-).
\]
Consequently, $\lz_{\xi_*}=\pi^{-1}(\gamma)$ lies in
$I^+_{\Hu}(\EuScript
S_-)\cup \EuScript S_-$, i.~e.\ in $U$.
Which means that $\Xi$ is closed. It is also non-empty and obviously
open. So,  $\Xi=[0,1]$.
\end{proof}

\begin{proof}[ of proposition \ref{nointe}. ]
Let $a,b\in D_M\cap M$ and let a curve $\varphi\subset D_M$ from $a$
to
$b$ be such that for some $x\in D$
\[
\varphi\subset I^+_D(x).
\]
Consider a homotopy $\lambda_\xi(\tau)$ with $\lambda_\xi\subset D$
being a geodesic segment from $x$ to $\varphi(\xi)$.
Suppose $\varphi\not\subset M$ and $p=\lambda_{\xi_p}(\tau_i)$,
$q=\lambda_{\xi_q}(\tau_j)$ are, respectively, the first and the
last points of $\varphi$ that do not lie in $M$, which means of
course that $\tau_i$ and $\tau_j$ are ($i$th and $j$th) roots of
$\lambda_{\xi_p}$ and $\lambda_{\xi_q}$, respectively.
By lemma~\ref{svyaz} the whole segment
$\lambda_{\xi_q}([\tau_i,\tau_j])$ belongs to $\A$. But a timelike
curve in $\A$ cannot intersect $\EuScript S_-$ more than once. So
actually $j=i$.

It follows then from corollary~\ref{traj} that there is a curve
$\mu_{\tau_i}\subset \EuScript S_-$ from $p$ to $q$ consisting of
points $\lambda_\xi(\tau)$ and thus lying in $D_M$ .
So, the curve composed of the segment of $\varphi$ from $a$ to $p$,
$\mu_{\tau_i}$, and the segment of $\varphi$ from $q$ to $b$ can
by a small variation (sending all inner points of this curve slightly
to the past) be transformed into a curve $\varphi'\subset D_M\cap M$.
\par
The connectedness of $D_M\cap\K$ and $D_M\cap\Ola$  is proved in the
same manner.
\end{proof}

%
%
\subsection{Geodesics intersecting the boundary of $\tilde H'$}
Let us introduce two new sets (see figure~\ref{mdiam}b):
$\Bu\equiv
\Bd_{\tilde H}\Out - \tilde \EuScript S_-$ and $\Bl\equiv \Bd_{\tilde
H}\Olt - \tilde \EuScript S_-$ (note that generally the surfaces
$\Blu $
may not be achronal in $\tilde H$). It is easy to see that $\Blu$ are
compact and that
\begin{equation}
\label{BlBu}\Bl \cap \Bu \subset \tilde\Theta.
\end{equation}

\par
Let $\lambda_\xi(\tau)$ with $\tau\geqslant 0,\,\xi\in[0,1]$ be such
a homotopy that $\lambda_\xi$ are
geodesic rays (not necessarily timelike) emitted from a single point
$p\equiv\lambda_\xi(0)\notin \overline{\tilde H'}$ and lying
in $\tilde H - \tilde \Theta$. Denote
by $h_{\wedge }(\xi)$ and $h_{\vee }(\xi)$ the intersections of
$\lambda_\xi$ with $\Bu$ and $\Bl$, respectively.  The corresponding
values of
$\tau$ we shall denote by $\tau_{\wedge }$ and $\tau_{\vee }$:
\[
\tau_{\wedge ,\vee}(\xi)\equiv \lambda_\xi^{-1}[h_{\wedge,\vee
}(\xi)].
\]
Since both $\tilde H$ and $\tilde H'$ are convex the segment of
$\lambda_\xi$ between $h_{\wedge }(\xi)$ and $h_{\vee }(\xi)$ (when
they both exist) lies in $\overline{\tilde H'}$. Moreover, as
$\lambda_\xi$ does not intersect $ \tilde\Theta$ it follows from
\eqref{BlBu} that the segment contains a point of
$\tilde H'$ and hence (see proposition~\ref{grvyp}) the whole segment
$\lambda_\xi[(\tau_{\wedge },\tau_{\vee })]$ lies in $\tilde H'$.
Which means, in particular, that $\lambda_\xi$ do not intersect $\Bu$
and $\Bl$ except in $h_{\wedge,\vee }(\xi)$.

\begin{teor}{Proposition}
\label{indl}
If $\lambda_{\xi_0}$ with  $\xi_0\in[0,1]$ intersects
both $\Bu$ and $\Bl$, then so do all the other $\lambda_{\xi}$ and
the functions $\tau_{\vee,\wedge}(\xi)$ are continuous.
\end{teor}
\begin{proof}
Consider the (non-empty by assumption) set $\Xi\subset [0,1]$ defined
by
\[
\Xi \equiv \{ \xi\colon\; \exists\, h_{\vee }(\xi),h_{\wedge }(\xi)\}
.\]
Suppose $\xi_1\in \Xi$.
Then
\begin{eqnarray*}
\lambda_{\xi_1}(\tau)\in \tilde\Hu,
\qquad\tau\in[\tau_\wedge-\epsilon,\tau_\wedge+\epsilon]\\
\lambda_{\xi_1}(\tau_\wedge-\epsilon)\in \tilde\Hu-
\overline{\Out},\qquad
\lambda_{\xi_1}(\tau_\wedge+\epsilon)\in \Out
\end{eqnarray*}
assuming that $\epsilon$ is sufficiently small and that (for
definiteness)
$\tau_{\wedge }<\tau_{\vee }$. Obviously this holds also for any
$\xi$
sufficiently close to $\xi_1$. Consequently all $\lambda_{\xi}$ with
such $\xi$ also intersect $\Bd{\Out}$ and hence (recall that
$\lambda_{\xi}$ do not pass through $\tilde \Theta$)
 $\Bu$. The same is true for $\Bl$.  So $\Xi$ is open.

Now let $\xi_2$ be a limit point of $\Xi$. $\Bu$ is compact, so it
contains the limit point $h$ of $h_{\wedge}(\xi)$:
\begin{equation}
\Bu\ni h=\lim_{\xi\to\xi_2}h_{\wedge}(\xi)=\lim_{\xi\to\xi_2}
\lambda_{\xi}(\tau_\wedge)=\lambda_{\xi_2}(\tau_*).
\label{pred}
\end{equation}
Similarly $\lambda_{\xi_2}$ must intersect $\Bl$.  So, $\Xi$ is
closed and hence is equal to $[0,1]$. From \eqref{pred} it follows
also that $\tau_{\wedge ,\vee}(\xi_2)=\tau_*=\lim_{\xi\to
\xi_2} \tau_{\wedge ,\vee}(\xi)$ whence $\tau_{\wedge,\vee }(\xi)$
are continuous.
\end{proof}

\begin{teor}{Remark}
\label{nakr2} As was mentioned in remark~\ref{nakr}, $N$ is an area in
a covering of $\tilde H - \tilde\Theta$.  Because this area is
bounded, not \emph{all} curve in $\tilde H -
\tilde\Theta$ can be
lifted to a continuous curve in $N$. It is easy to check, however,
that
if a curve $\tilde \gamma\colon\:[0,1]\to \tilde H$ satisfies the
following condition
\begin{equation}
    \label{clas} \tilde \gamma(0)
    \notin \overline{\tilde H'},\qquad
    \tilde \gamma\cap\Bu = \varnothing,\ftext{{} or } \tilde
\gamma\cap\Bl =
    \varnothing ,
    \end{equation}
then there is a unique curve $\gamma \subset N$ connecting $\pi^{-
1}(\tilde\gamma(0))$ with $\pi^{-1}(\tilde\gamma(0))$ and
satisfying $\pi( \gamma)=\tilde\gamma$. It is this  $\gamma$ that we
understand by $\pi^{-1}(\tilde\gamma)$ from now on.
\end{teor}

\begin{teor}{Proposition}
\label{postcl}
Let $\lambda_\xi(\tau)$ be the homotopy from proposition~\ref{indl},
$\tilde\mu(\xi)\equiv{\lambda}_\xi(1)$, and
${\tilde\gamma}_\xi(\tau)\equiv\ogr{\lambda_\xi(\tau)}{\tau\in[0,
1]}$. Suppose that $\tilde\mu$ does not
intersect $\tilde\EuScript S_-$ and that $\tilde\mu(0) \notin
\overline{\tilde H'}$. If
${\tilde\gamma}_0$
satisfies \eqref{clas} then so does ${\tilde\gamma}_1$.
\end{teor}
\begin{proof}
Suppose for contradiction that ${\tilde\gamma}_1$ (and hence
$\lambda_1$) intersects both $\Bu$ and $\Bl$:
    \[
    \tau_{\wedge }(1),\tau_{\vee }(1) \leqslant 1
    .\]
Then (by proposition~\ref{indl}) so does
$\lambda_0$ \emph{in contrast to}
${\tilde\gamma}_0$, which is only possible if $\lambda_0(1)$ lies on
$\lambda_0$
closer to $p$ than $h_{\wedge }(0)$ (and thus also than $h_{\vee
}(0)$ assuming as before that $\tau_{\wedge }<\tau_{\vee }$). So,
\[
    \tau_{\wedge }(0), \tau_\vee(0) > 1
\]
and there must exist $\xi'$ and $\xi''$:
$$
\xi'\equiv\max\{\xi|\;\tau_{\wedge }(\xi)\geqslant 1\},\quad
\xi''\equiv\min\{\xi|\;\xi>\xi',\,\tau_{\vee}(\xi)\leqslant 1\},\quad
0<\xi'<\xi ''<1.
$$
The (open) segment of $\tilde\mu$ between $\xi'$ and $\xi''$ is a
continuous curve lying within $\tilde H'$ (since $\tau_{\wedge
}< 1<\tau_{\vee }$ here) and connecting $\Out$ with
$\Olt$ without
intersecting $\tilde\EuScript S_-$, which is impossible.
\end{proof}

%
\section{Proof of the theorem }
\label{dvo}
For any region $U$ denote by $\Z{U}$ the \emph{causality violating
subset} of $U$, that is the set of all points $p$ satisfying
$J^+_U(p)\cap J^-_U(p)\neq p$. Note that $\Z{U}$ is determined by the
causal structure of $U$, but not by that of the ambient spacetime
$U'\supset U$ (if it exists), in the sense that generally
$\Z{U}\varsubsetneq\Z{U'}\cap U$.

\begin{teor}{Proposition}
\label{exte}
Any extendible LCC spacetime $M$ has an LCC extension whose causality
violating subset is $\Z{M}$.
\end{teor}
\begin{proof}
We shall prove that an extension $\A$, which (as was shown in
section~\ref{Constr}) can be built for any extendible LCC $M$,
satisfies the requirement of the proposition, i.~e.\ that $\A$ is LCC
and $\Z{\A}=\Z{M}$. The latter follows immediately from remark~\ref{I}
(since $\A$ can be isometrically immersed into $M_\vartriangle$)
 and to prove the former suppose that
the assertion is false. Then there exists a future-directed curve
$\lambda\subset D$ from $a$ to $b$ such that
\[
a,b\in  D_M,\quad  \lambda\not\subset D_M.
\]
Without loss of generality we may assume that $\lambda$ does not
intersect $\EuScript S_\Bumpeq$ (and thus has no roots). By
proposition~\ref{kriv} this implies that the ends of $\lambda$ lie
either
both in $M$, or both in $\Ola$, or both in ${\K}$. But $M$ and
$H'_{\mathcal N}$
are LCC, which implies that the ends of $\lambda$ cannot lie in a
connected component of $D\cap M$, or $D\cap H'_{\mathcal N}$ and
thus by proposition~\ref{nointe} in $D_M\cap M$, or $D_M\cap
H'_{\mathcal N}$. So, $a,b\in\K$.

Choose $a$ and $b$ so that $\pi(a), \pi(b)\notin \overline{\tilde
H'}$ (it always can be done). We want to show that they can be
connected by a \emph{timelike} curve --- it will be the geodesic
$\tilde\gamma_1$ --- lying in $D\cap \K$.
 To this end let us, first, pick a curve $\mu(\xi)\colon\;[0,1]\to
 D\cap \K$ connecting $a$ and $b$ (its existence is guaranteed by
proposition~\ref{nointe}).
 $\mu$ can always be chosen (cf.\
proposition~\ref{kriv}) so that for some $x,y\in D$
\[
\mu\subset\lr[D]{x,y}.
\]
Consider
the geodesics $\tilde \gamma_\xi\subset \tilde H$ with
$\tilde \gamma_\xi(0)=\tilde \mu(0)$ and $\tilde \gamma_\xi(1)=\tilde
\mu(\xi)$ (as usual $\tilde \mu\equiv\pi( \mu)$). Let $\Xi$ be the
set of all $\xi$ such that
\begin{equation}
\label{des}
\tilde \gamma_{\xi'}\ftext{{} satisfies \eqref{clas},}\quad
\gamma_{\xi'}\equiv\pi^{-1}(\tilde \gamma_{\xi'})
\subset\lr[D]{x,y}
\qquad\forall\xi'\leqslant\xi.
\end{equation}
$\Xi$ is evidently open and (as small $\xi$ obviously lie in it)
non-empty. To see that it is also closed consider its limit point
$\xi_c$
\[
\xi_n\to\xi_c,\qquad\xi_n\in\Xi.
\]
By construction none of $\tilde \gamma_{\xi'}$ with $\xi'<\xi_c$
 meets $\tilde\Theta$.
It is easy to check
 that neither does $\tilde \gamma_{\xi_c}$. Whence by
 proposition~\ref{postcl} the condition~\eqref{clas} holds for
$\xi_c$. Consequently there exists a geodesic
$\gamma_{\xi_c}\equiv\pi^{-1}(\tilde \gamma_{\xi'})$, which
(being the limit of $\gamma_{\xi_n}$) lies in $\overline{\lr[D]{x,y}}
\subset D$ and moreover (since  $a$ and $b$ belong to a convex subset
$\lr[D]{x,y}$ of $D$) in
$\lr[D]{x,y}$. Thus the whole condition~\eqref{des} holds for $\xi_c$
and $\Xi$ is closed.  So, $\Xi=[0,1]$.\par

We have proved, thus, that $\gamma_1\subset D\cap N$. Besides,
$\gamma_1$ is
timelike (since $b$ is connected to $a$
with the timelike $\lambda\subset D$) and $n_\pm[\gamma_1]=0$ (since
$n_\pm[\lambda]=0$), whence $\gamma_1\subset \K$. So, in $D\cap \K$
a timelike curve from $a$ to $b$ exists indeed. Which yields
\begin{subequations}
\label{DK}
\begin{eqnarray}
\label{DKa}\lambda(\epsilon),\lambda(1-\epsilon)
\in D_K, \\
\label{DKa'}\lambda([\epsilon,1-\epsilon])\subset D_1,\quad
\lambda([\epsilon,1-\epsilon])\not\subset D_K,\\
D_1\cap\EuScript
S_\Bumpeq=\varnothing,
\label{DKb}
\end{eqnarray}
\end{subequations}%
where $D_1\equiv\lr[D]{a,b}$, $D_K$ is a component of $D_1 \cap \K$,
and $\epsilon$ is chosen appropriately small. To see that \eqref{DK}
is, in fact, impossible consider the region $D_1
\cup \K$ of the spacetime $M^e$ as a separate spacetime and extend it
to a spacetime $H^e$ by gluing $\tilde H$ to it [\eqref{DKb} ensures
 that the resulting manifold is Hausdorff):
\[
H^e\equiv (D_1 \cup \K)\cup_{\varpi_K} \tilde H
\]
(in other words $H^e$ is what results when the `hole' in $\K$, which
appeared when $D_1 \cup \K$ was cut out of $M^e$, is `closed up' by
gluing its edges (the former $\EuScript S_-$ and $\EuScript S_+$ )
together). $H^e$ is an
extension of  $\tilde H$ and $D_1$ is (see remark~\ref{nasl1})  its
diamond subset. So
 (\ref{DK}a,b) contradict the local causal convexity of $\tilde H$.
\end{proof}
\begin{teor}{Proposition}
\label{max}
Any  extendible LCC spacetime $M$ has a maximal  extension  $\zv M$
with $\Z{\zv M}=\Z{M}$.
\end{teor}
\begin{proof}
Consider the set  $\mathsf E$ of all pairs $(V,\zeta)$, where
$\zeta$
is an isometric imbedding of
$ M$ into $V$ (we normally shall not distinguish $\zeta(M)$ and $M$)
and
$V$ is an LCC extension of $M$ with $\Z{V}=\Z{M}$. Pick a point $p\in M$ and
introduce the following order relation in $\mathsf E$ (cf.\
\cite{HawEl}).
We write $(V_1,\zeta_1)\leqslant
(V_2,\zeta_2)$ if there exists an isometric imbedding
 $\vartheta_{12}\colon\:V_1\to V_2$  satisfying the
conditions:
\[
{\zeta_2}^{-1}\circ\vartheta_{12}\circ\zeta_1(p)=p,\quad
\ogr{\rmd({\zeta_2}^{-
1}\circ\vartheta_{12}\circ\zeta_1)}{p}=\mathrm{id}
.\]
It is easy to check that in agreement with its notation $\leqslant$
is a partial order (which it would not be if we relax any of the
two conditions). Let $\{(V_\alpha,\zeta_\alpha)\}$ be a chain with
respect to $\leqslant$ and let
\[
V = \big(\bigcup_\alpha V_\alpha\big)\mathop{\backslash}{\sim},
\]
where the equivalence is defined as follows:
\[
x\sim y \; \Leftrightarrow\; \exists\, \alpha_1,\alpha_2\colon\:
\vartheta_{\alpha_1\alpha_2}(x)=y,\ftext{{} or
}\vartheta_{\alpha_1\alpha_2}(y)=x.
\]
Let us further introduce the following notations:  $V^e$
is an extension of $V$; $D$ is a diamond subset of $V^e$; $D_V$ is a
connected component of $V^e\cap D$; $\varphi\subset D_V$ is a curve
between points $a$ and $b$; $\lambda_{ab}\subset D$ is a timelike
curve
from $a$ to $b$; and $\ell\subset V$ is a closed causal curve.
\par
Since the curves $\varphi$ and $\ell$ are compact we can
choose a finite number of $V_\alpha$: $\{V_{\alpha_n}\},\;n=1,\dots
N$ so that (from now on we do not distinguish $V_{\alpha}$ and their
images in $V$)
    \begin{equation}
    \label{cau}
    \varphi,\ell\subset
    \bigcup_n V_{\alpha_n}\subset V_{\alpha_0}.
    \end{equation}
Along with the whole $\varphi$ the points $a$,$b$ lie in a connected
component of $V_{\alpha_0}\cap D$ and (since $V_{\alpha_0}$ is LCC) so
does $\lambda_{ab}$. Hence $V$ is LCC. Also $\Z{V}$ is equal to
$\Z{M}$, since otherwise [by \eqref{cau}] $\Z{V_{\alpha_0}}$ would not
be. Finally, \C\ holds in $V$ since it is covered by the open sets $
V_{\alpha_n}$ where it does.
So, $(V,\zeta_1)\in \mathsf E$ and obviously $(V,\zeta_1)=\sup
\{(V_\alpha,\zeta_\alpha)\}$.
Which means that $\leqslant$ is an inductive order in $\mathsf E$. It
follows then by the Zorn lemma that $M$ has an LCC extension $\zv M$
with $\Z{\zv M}=\Z{M}$ which cannot be extended to any larger
spacetime with such properties. By proposition~\ref{exte} $\zv M$ is
maximal.
\end{proof}

\par\vspace{\topsep}
\begin{proof}[ of the theorem. ]
Consider the set of all possible extensions of $U$ of the form $I^-
_{U'}(U)$,
where $U'$ is an arbitrary spacetime. By repeating the procedure
employed in
proposition~\ref{max} it can be shown that $U$ can be imbedded into
a maximal element --- let us call it $M$ --- of that set. $M=
I_{M^e}^-(U)$ in any  extension $M^e$ of $M$. So $M$ is
obviously LCC (since a past-directed curve even
cannot leave $M$) and (tautologically) $\Z{M}\subset I_{M^e}^-(U)$.
Which being combined with  proposition~\ref{max} proves the theorem.
\end{proof}

\ack
I wish to thank R. R. Zapatrin for useful discussion and Starlab NV
for wonderful conditions which I enjoyed while working on this paper.

\end{document}


\title{Corrigendum on: No time machines in classical general
relativity}
\author{S Krasnikov}%
\address{Central Astronomical Observatory at Pulkovo,
St.Petersburg, 196140, Russia}
\ead{Gennady.Krasnikov@pobox.spbu.ru}

\section*{}
The theorem formulated in \cite{eotm} reads (after some rewording):
\begin{teor*}{Theorem} 1) Any  spacetime $U$ has a  maximal   extension
$\zv M$ such that all closed causal curves in $\zv M$ (if they exist
there) are confined to the chronological past of $U$. 2) The assertion remains true, even if the definition of spacetime is complemented by an arbitrary local geometric  condition \C.
\end{teor*}

Obviously the meaning and the validity of the second part of the theorem depends crucially on what is understood by ``local" and ``geometric". The latter term is transparent---a condition is called geometric, iff it holds in a spacetime $M$, when and only when it holds in any spacetime isometric to $M$---but the situation with locality is more subtle. In \cite{eotm} the following inadequate definition crept into the text:
\begin{teor}{Definition}\label{def:locM}
We call a condition \C\ \emph{local} if the following is true:
\C\  holds in a spacetime $M$ if and only if it holds in any $U$
which is isometric to an open
subset of $M$.
\end{teor}
The disadvantage of this definition is that it does not guarantee the validity of the implication
\[
A,B \in  \Se    \quad\Rightarrow\quad \bigl(A\cup B\bigr) \in  \Se  ,
\]
where by $\Se $ we have denoted the set of all spacetimes satisfying the condition \C. This has two consequences, as is discussed in \cite{ManchakLoc,LowZorn}:\\
i). The definition does not properly  capture the notion of locality;\\
ii). The  second part of the theorem is erroneous, see \cite{ManchakLoc} for a counterexample. The error in the proof is the false implication that
$$
    \bigl[M\in \Se \mbox{ and } N \mbox{ and is locally isometric to } M\bigr]   \quad\Rightarrow\quad N\in \Se
\eqno{(*)}
$$
(note that the validity of $(*)$ is the \emph{only} property of \C\ used in proving the theorem).

Now consider an alternative
 \begin{teor}{Definition}\label{def:loc}
   The  condition (property) \C\
 is \emph{local}, if for any open covering $\{V_\alpha\}$ of an arbitrary spacetime $M$ the following equivalence is true\footnote{Cf. ``particularly interesting conditions" imposed on properties in Appendix B of \cite{Ger2}.}
\[
M\in\Se \qquad\Leftrightarrow\qquad V_\alpha\in\Se \quad\forall\,\alpha.
\]
\end{teor}
  The two definitions are not equivalent. For example,  the former is satisfied by the---non-local according to definition~\ref{def:loc}---property ``to have timelike  diameter\footnote{See \cite{BeE} for definition.}
   not greater than 1". Definition~\ref{def:loc} is closer to  the intuitive notion of locality: in particular,   it   makes the implication~($*$) true, which solves all the problems mentioned above.

   Thus, the only correction needed by \cite{eotm} is the replacement of definition~\ref{def:locM} by definition~\ref{def:loc}.